\documentclass{article}
\usepackage[utf8]{inputenc}
\usepackage{graphicx, amsmath}
\usepackage[left=2cm, right=2cm, top=2cm]{geometry}
\usepackage{graphicx}
\usepackage{tabularx}
\usepackage{tabu}
\usepackage{longtable}
\usepackage[colorlinks]{hyperref}
\usepackage{listings}
\usepackage{textgreek}%For greek letters outside math mode

\usepackage[minbibnames=3,natbib=true, sorting=none]{biblatex}
\usepackage{authblk}

\begin{document}

\author[1]{James Williams}
\author[1,3,4]{Rajveer Nehra}
\author[2]{Elina Sendonaris}
\author[1]{Luis Ledezma}
\author[1]{Robert M. Gray}
\author[1]{Ryoto Sekine}
\author[1]{Alireza Marandi} 
\affil[1]{Department of Electrical Engineering, California Institute of Technology, Pasadena, California 91125, USA}
\affil[2]{Department of Applied Physics, California Institute of Technology, Pasadena, California 91125, USA}
\affil[3]{Department of Electrical and Computer Engineering, University of Massachusetts Amherst, Amherst, Massachusetts, 01003, USA}
\affil[4]{Department of Physics, University of Massachusetts Amherst, Amherst, Massachusetts, 01003, USA}
\title{Ultra-Short Pulse Biphoton Source in Lithium Niobate Nanophotonics at 2 \textmu m}

\maketitle

\section{Abstract}
Photonics offers unique capabilities for quantum information processing (QIP) such as room-temperature operation, the scalability of nanophotonics, and access to ultrabroad bandwidths and consequently ultrafast operation. Ultrashort-pulse sources of quantum states in nanophotonics are an important building block for achieving scalable ultrafast QIP, however, their demonstrations so far have been sparse. Here, we demonstrate a femtosecond biphoton source in dispersion-engineered periodically poled lithium niobate nanophotonics. We measure 17 THz of bandwidth for the source centered at 2.09 \textmu m, corresponding to a few optical cycles, with a brightness of 8.8 GHz/mW. Our results open new paths towards realization of ultrafast nanophotonic QIP.
\section{Introduction}

\begin{figure*}
\includegraphics[width=\textwidth]{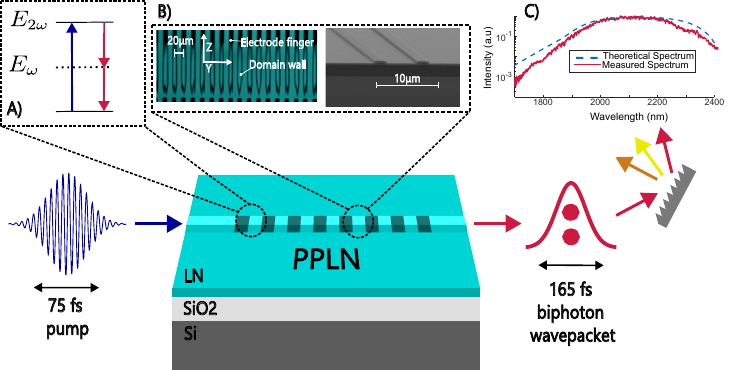}
\caption{A diagram of the source operation. Femtosecond pump pulses are injected into a dispersion-engineered periodically polled lithium niobate (PPLN) waveguide to produce an ultra-fast photon pair via type-0 spontaneous parametric down-conversion (SPDC). A) An illistration of the spontaneous parametric down converison (SPDC) process used to generate photon pairs. B) 2-photon microscopy image of the polled region and an SEM of a representative device. C) The theoretical and measured SPDC spectra.}
\label{fig:fig1}
\end{figure*}

Since the first observations of the quantum nature of light, considerable efforts have gone into building systems for gaining a quantum advantage over purely classical technologies to benefit a variety of fields including sensing, metrology, communication, and computation\cite{degen2017quantum, gisin2007quantum, horowitz2019quantum}. While many physical platforms have been studied to exploit quantum phenomenon, photonics stands out as it offers several benefits including room temperature operation and scalability. Integrated photonics \cite{siew2021review, zhu2021integrated} allows the realization of many devices in a chipscale footprint, paving the way for the development of large-scale integrated systems for QIP at room temperature \cite{wang2020integrated, pelucchi2022potential}. 

In recent years, thin-film lithium niobate (TFLN) has emerged as a favorable platform for the development of entangled photon sources given its high $\chi^{(2)}$. Additionally, dispersion engineering enables the generation of broadband entangled photon pairs with short coherence times. Broadband operation is important in a variety of quantum applications as it leads to improved sensitivity in optical coherence tomography \cite{abouraddy2002quantum}, greater spectral coverage for quantum spectroscopy \cite{hickmansinglephoton, whittaker2017absorption}, sharp temporal behavior for atomic state control \cite{dayan2004two}, and lithographic resolution exceeding the classical diffraction limit \cite{pavel2014quantum}. Wavelength-multiplexing schemes \cite{wengerowsky2018entanglement, pseiner2021experimental} as well as high-dimensional entanglement protocols from quantum networks \cite{wei2022towards, wengerowsky2018entanglement} can also make good use of broadband sources.

There have been many successful demonstrations of broadband biphoton sources in TFLN \cite{javid2021ultrabroadband, zhang2023scalable, xue2021ultrabright}. However, these devices are still limited to the telecom wavelength range. Moreover, they have been demonstrated in the continuous-wave (CW) regime and have been unsuitable for ultrafast operation given their dispersive properties. Operation in the ultrafast regime requires near-zero group velocity dispersion (GVD) conditions for both the pump and signal wavelengths, as well as near-zero group velocity mismatch (GVM) between the pump and signal. Ultrashort-pulse sources of quantum states of light are particularly important because of opportunities in time-division multiplexing and the enhancement of nonlinear interactions as a result of increased peak power. By time-division multiplexing with well-defined nanosecond time bins, large-scale quantum states are realized using fiber-based systems \cite{asavanant2019generation, larsen2019deterministic}. Their extension to the ultrafast femtosecond time bins can enable the generation and manipulation of similarly large-scale quantum states in integrated photonics \cite{gray202340}. Stronger nonlinear interactions as a result of temporal confinement are also an important step towards all-optical non-Gaussian sates and operations \cite{yanagimoto2023quantum, onodera2022nonlinear}. While our implementation currently relies on a table-top mode-locked laser, significant progress has been made in integrating short pulse sources on chip, especially in LN nanophotonics, with a 4.8-ps 10-GHz chip-scale mode-locked laser demonstrated in \cite{guo2023ultrafast} and a chip-scale electro-optic comb source with a 520-fs duration and a 30-GHz repetition rate demonstrated in \cite{yu2022integrated}.

Dispersion engineering in TFLN using the waveguide geometry provides new opportunities for classical and quantum ultrafast photonics beyond broadband CW operation. Particularly, achieving near-zero group-velocity mismatch between the signal and pump along with zero group-velocity dispersion at the signal and pump wavelengths simultaneously has enabled record-breaking gain-bandwidth products \cite{ledezma2022intense}, ultralow-energy ultrafast all-optical switching \cite{guo2022femtojoule} and the on-chip generation and measurement of broadband squeezed states \cite{nehra2022few}.  Unlike the previous nanophotonic biphoton sources \cite{steiner2021ultrabright, zhao2020high, elkus2019generation, javid2021ultrabroadband, guo2017parametric, prabhakar2020two, signorini2021silicon, sanna2022integrated, kumar2021mid, rosenfeld2020mid}, we utilize a near-zero dispersion regime for the realization of ultrashort-pulse photon pairs in nanophotonics. Operating in the 2-\textmu m band in lithium niobate is particularly advantageous as it yields better fabrication tolerances compared with similar geometries for the 1550-nm band \cite{kuo2022towards}, and low GVD for both pump and signal simultaneously as well as the ability to match their group velocities and create strong temporal confinement for generated photon pairs \cite{jankowski2021dispersion}.

Biphoton sources beyond the telecommunication band are also important for quantum key distribution (QKD) applications. For instance, recent work on free-space QKD at 1550 nm has made a compelling case for moving to longer wavelengths to avoid limitations caused by solar irradiance \cite{liao2017long}, which decreases at longer wavelengths \cite{solartables}. Integrated photonics platforms can also benefit from operation at longer wavelengths because of lower scattering losses \cite{hagan2016mechanisms} and more efficient high-speed modulators \cite{cao2018high, sadiq201640}. Classical and quantum networks are becoming practical in the 2-\textmu m window given the progress on sources \cite{phelan2018high, yang2019high}, as well as low-dispersion and low-nonlinearity photonic crystal fibers \cite{agrell2016roadmap, richardson2016new}, and high-speed detectors \cite{ye2015alingaas}. Along with thulium/holmium amplifiers offering 40 THz of gain bandwidth \cite{li2013thulium}, these technologies can facilitate the construction of wide area classical and quantum networks in the mid-IR. Quantum effects and secure communications have also been successfully demonstrated in the 2-\textmu m band with Hong-Ou-Mandel visibility of 88.1\% measured in bulk PPLN by \cite{prabhakar2020two} and polarization-based quantum key distribution implemented by \cite{dada2021near} with a key rate of 0.254 bits/pair.

In this work, we present the first ultrafast biphoton source in nanophotonic PPLN capable of supporting ultra-short pulse propagation at both the pump and signal/idler wavelengths (Fig.\ref{fig:fig1}). Our source covers 17 THz of 3-dB bandwidth centered at 2.09 \textmu m, and produces a 165-fs biphoton wavepacket at the output. Through 2 and 3-fold coincidence counting experiments, we demonstrate a coincidence-to-accidentals ratio (CAR) of 945, a pair generation rate of 8.8 GHz/mW or 440 kHz/mW/GHz, and a heralded $g_H^{(2)}(0)$ of 0.027, all of which are state-of-the-art for this wavelength range in nanophotonics. Combined with the recent advances in ultrafast lithium niobate nanophotonics \cite{ledezma2022intense, nehra2022few, guo2022femtojoule, yu2022integrated}, and the wide variety of high-performance components \cite{zhu2021integrated}, our source demonstrates a practical path towards ultrafast on-chip QIP.

\begin{figure}

\centering\includegraphics[width=6.6cm]{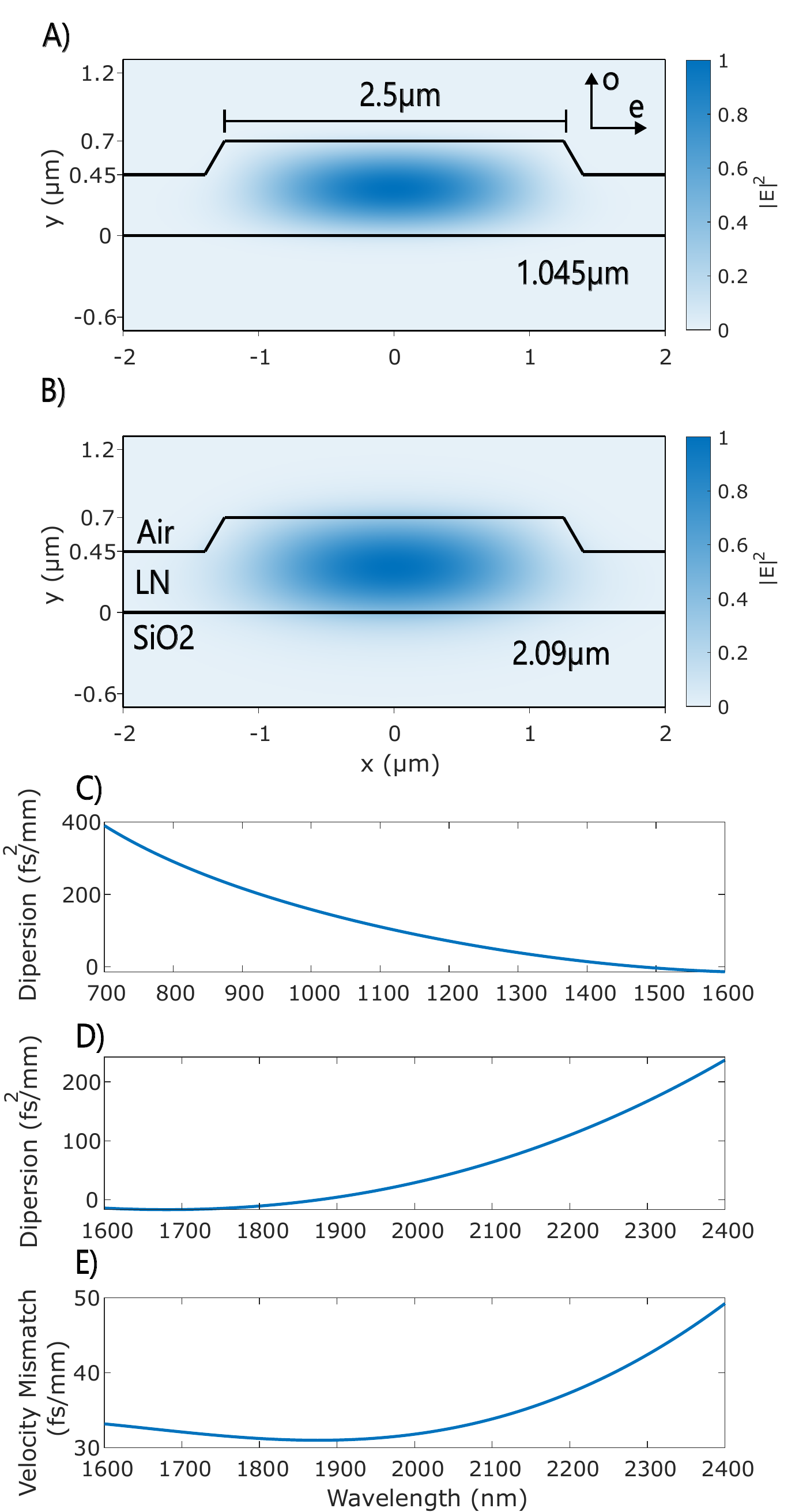}
\caption{A) The quasi-TE waveguide mode for pump light at 1.045 \textmu m. Arrows at the top right denote the ordinary and extraordinary material axes. B) The quasi-TE waveguide mode for signal light at 2.09 \textmu m. C) Dispersion profile for the pump wavelength. D) Dispersion profile for the signal wavelength. E) Group velocity mismatch at different signal wavelengths relative to 1.045 \textmu m.}
\label{fig:mode_dispersion}
\end{figure}

\begin{figure*}
\centering\includegraphics[width=15cm]{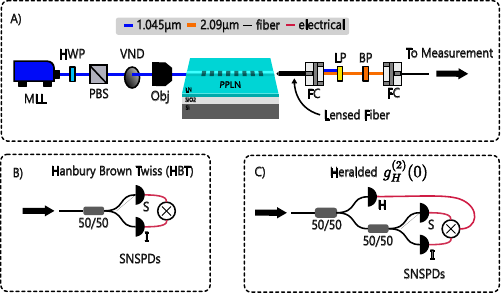}
\caption{A) The experimental setup used to characterize the source. MLL is a 250-MHz 75-fs mode-locked laser centered at 1.045 \textmu m. HWP is a half-wave plate. PBS is a polarizing beam splitter. VND is a variable neutral density filter. Obj is a reflective objective. FC is a freespace to fiber coupler. LP is a low-pass filter. BP is a band-pass filter. SNSPDs are superconducting nanowire single photon detectors. 50/50 are balanced fiber beamsplitters. H, S, and I denote the heralding, signal, and idler channels respectively. B) Measurement setup for performing the Hanbury Brown-Twiss (HBT) experiment. C) Measurement setup for determining $g_H^{(2)}(0)$. The SEM image inset is of a representative device \cite{ledezma2022intense}.}
\label{fig:exp_setup}
\end{figure*}

\section{Device Design and Fabrication}

Our source is designed for type-0 degenerate spontaneous parametric down-conversion (SPDC) while also optimizing the dispersion parameters for short-pulse propagation. To enjoy the benefits of short pulse operation, the group-velocity dispersion at both the pump and signal wavelengths must be minimized to maintain the temporal confinement of the pulses as they propagate along the length of the waveguide. Our waveguide design also minimizes the group-velocity mismatch between the pump and signal/idler wavelengths to minimize the temporal interaction window of the pump and generated photon pairs \cite{vanselow2019ultra}, allowing for the creation of an ultrashort signal/idler wavepacket. This short wavepacket can facilitate strong nonlinear interactions given the peak power enhancement and enable the use of time bins shorter than 1 ps for clock speeds exceeding 1 THz. We target our waveguide design for conversion from a pump centered at 1.045 \textmu m to a degenerate signal/idler pair at 2.09 \textmu m. The dispersion curves for our design are presented in Fig.\ref{fig:mode_dispersion}. We achieve a GVD of 135 $fs^2/mm$ and 60 $fs^2/mm$ for the pump and signal/idler respectively, and a GVM of 33 $fs/mm$. For comparison, bulk lithium niobate has has a GVD of 246 $fs^2/mm$ for pump light, a GVD of -56 $fs^2/mm$ for signal/idler photons, and a GVM of 115 $fs/mm$.

Our near-zero GVM and GVD regime of operation leads to a broad spectrum of signal/idler pairs as shown in Fig.\ref{fig:fig1} with a 3-dB bandwidth of 17 THz. For a 5-mm long device, we achieve a maximum temporal length for the signal/idler wavepacket of 165 fs. The duration of this wavepacket can be estimated from the temporal overlap of the pump pulse with the vacuum modes into which the waveguide phase-matching and dispersion permit SPDC. As the pump propagates down the waveguide and walks off in time from the 2 \textmu m vacuum field, an large number of time-delayed vacuum modes within the walk-off window can experience the creation of a photon pair. The total wavepacket duration is approximated from the pump pulse duration and the total walk-off time resulting from the GVM using Eq.27(a) from \cite{wasilewski2006pulsed}. Shorter gain windows, the gain being responsible for pair generation, for similar devices have been demonstrated in \cite{li2022all}. The temporal length of the photon pair can be directly measured using techniques developed in \cite{sensarn2010generation} and \cite{chekhova2018broadband}. Substantial GVD at either the pump or signal/idler frequencies will also temporally broaden generated photon pairs either by increasing the effective temporal gain window or by dispersing photon pairs after their creation \cite{roeder2023measurement}. For 700-nm thick lithium-niobate on insulator (LNOI), achieving a large enough GVD necessary for the GVD to be the primary contribution to the temporal length of the photon pairs is generally difficult without resorting to extremely narrow waveguide geometries or deep etch depths. Therefore, our focus is on minimizing the GVM as this is the dominant contribution from the waveguide geometry to our temporal length. There exist other geometries for LNOI that also experience low GVM and GVD thanks to the relaxed fabrication tolerances and flatter dispersion curves around 1 and 2 \textmu m \cite{jankowski2021dispersion, ledezma2022intense}.

We calculate the waveguide dispersion by first measuring the width and etch depth via atomic force microscopy, and then simulating the exact geometry with a mode-solver to find the effective refractive index, group velocity, and second-order dispersion. In addition to temporal confinement, the waveguide geometry spatially confines both the pump and signal/idler modes to the fundamental quasi-TE mode, providing a large mode overlap to increase the pair generation rate. The mode profiles are plotted in Fig.\ref{fig:mode_dispersion}A-B.

While not the focus of our work, many recent works on photon-pair sources focus on using dispersion or poling domain engineering to minimize the number of temporal field-orthogonal modes \cite{brecht2015photon} present in the output pairs \cite{xin2022spectrally, dixon2013spectral, zhong2020quantum, mosley2008heralded, graffitti2018independent}. The structure and occupancy of these modes can be found by performing a Bloch-Messiah decomposition on the joint-spectral intensity (JSI) of the signal and idler photons \cite{houde2023waveguided}. The JSI itself is estimated by the product of the energy conservation of the pump and the phase matching of the waveguide as a function of the signal and idler frequencies. A more detailed discussion of this process is presented in the supplementary. In the context of SPDC, minimization of the mode number is equivalent to removing the spectral correlations from the photon pairs. This creates indistinguishable photons, which are especially important for applications where multiple independent biphoton sources are combined \cite{kok2007linear, humphreys2013linear, briegel1998quantum, spring2013boson} as this indistinguishably leads to a stronger interference of single photons from different sources. Single-mode operation for the degenerate type-0 case without the use of filtering has remained elusive due to constraints stemming from the shape of the phase matching function as well as an inherent inability to satisfy the group velocity criteria used in \cite{grice2001eliminating, xin2022spectrally}. We estimate a mode number of 16 from the decomposition of the JSI. For an ideal experiment with perfectly achromatic optics and detectors, this mode number will not limit the performance in CAR, generation rate, or heralded $g_H^{(2)}(0)$ of our source. In practice, the SNSPDs used for single photon detection experience a sharp fall-off in efficiency for increasing wavelength as does the transmissivity of the SMF-28 fibers leading up to them. We therefore perform all SNSPD measurements through a 48-nm bandpass filter centered at 2.09 \textmu m to limit the impact of achromatic detection on our measurements. This reduces our mode number to 3. Using this filter during measurement prevents us from directly observing the total brightness of the source. Therefore, using a technique similar to \cite{steiner2021ultrabright}, we calculate the total brightness by computing the ratio of the measured SPDC spectrum and the filter pass band. Multiplying the measured brightness by this ratio gives the total brightness.

We fabricate our devices using commercially available thin-film lithium niobate on silica wafers from NANOLN, with a 700-nm thick X-cut thin-film on 4.7-\textmu m thick SiO2. We start by using electron beam lithography to pattern poling electrodes deposited via metal evaporation. After poling, we lithographically define and etch waveguides with dry Ar+ plasma. The inset of Fig.\ref{fig:fig1}B shows a 2-photon microscopy image of the poled region before etching as well as an SEM image of the waveguide facet at the edge of the chip.

\section{Experimental Setup}

The experimental setup used to characterize our biphoton source is shown in Fig.\ref{fig:exp_setup}. A 75-fs 250-MHz mode-locked laser centered at 1.045 \textmu m is directed through a motorized half-wave plate (HWP) and polarizing beam splitter (PBS) combination used to digitally control the input power to the experiment. Laser light is then collected and focused onto the input facet of the waveguide via a reflective objective to minimize dispersion. Inside the waveguide, photons from the pump are split into signal/idler pairs via degenerate type-0 SPDC. Light from the output facet is collected with a lensed PM-2000 fiber from OZ Optics. This fiber is directed into freespace with a reflective coupler and passed through one 1330-nm long pass filter, three 1500-nm long pass filters, and a 48-nm bandpass filter centered at 2.09 \textmu m to provide 150 dB of pump rejection. The photon pairs are then coupled back into fiber and passed through a series of fiber-coupled 50/50 splitters to perform Hanbury Brown-Twiss (HBT) and heralded $g_H^{(2)}(0)$  experiments for measuring the CAR, pair generation rate, and single photon behavior. 

\begin{figure}[ht!]

\centering \includegraphics[width=7cm]{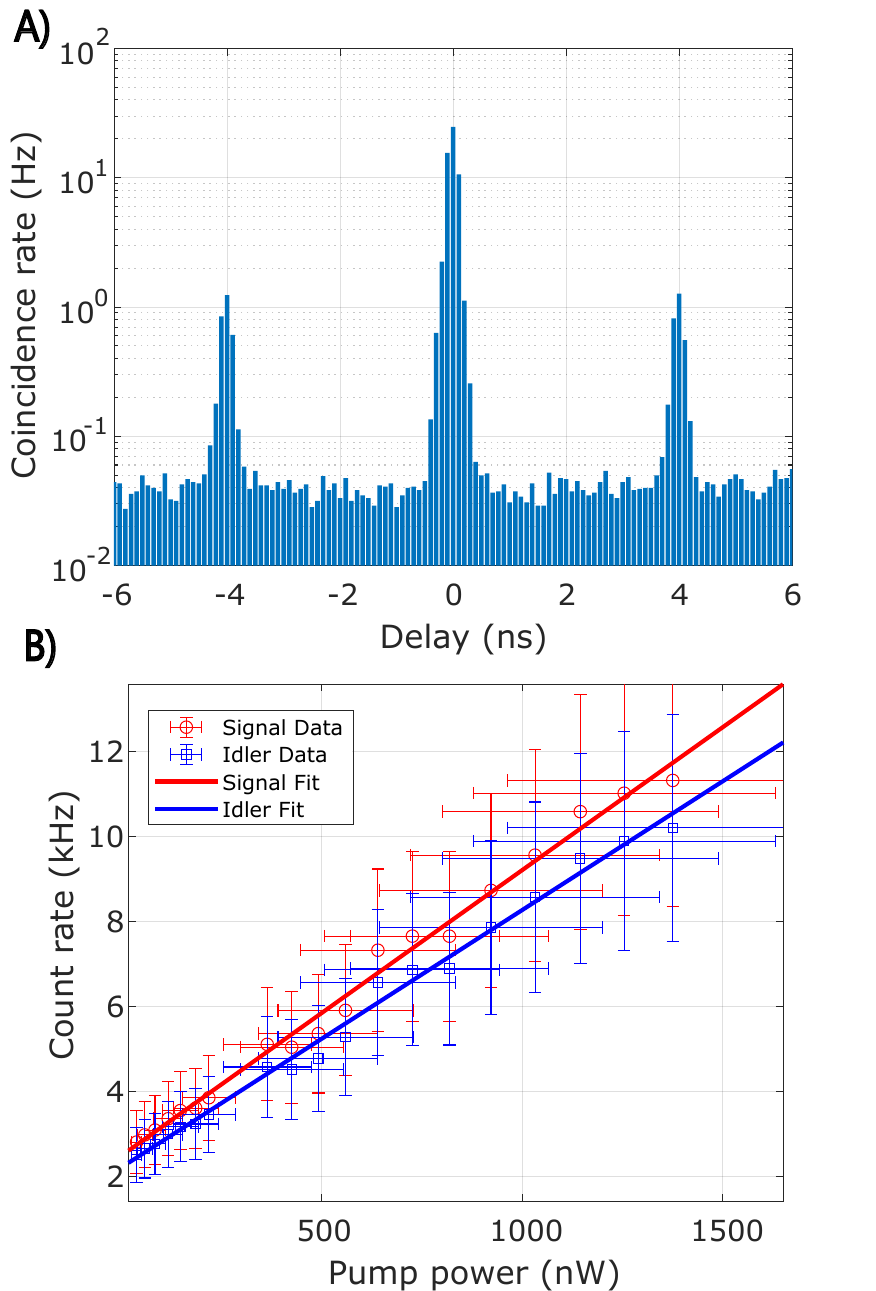}
\caption{A) Example histogram from an HBT experiment taken at 11 \textmu W of on-chip pump power. The main coincidence peak, caused by the detection of a signal/idler pair from the same generation event, is centered at 0-ns delay while accidentals peaks, caused by signal and idler photons created from consecutive and hence uncorrelated generation events, are present to the left and right at $\pm$4 ns. B) Count rates for the signal and idler channels compared with the fitted linear model.}
\label{fig:raw_hbt}
\end{figure}

\begin{figure}[ht!]

\centering \includegraphics[width=6cm]{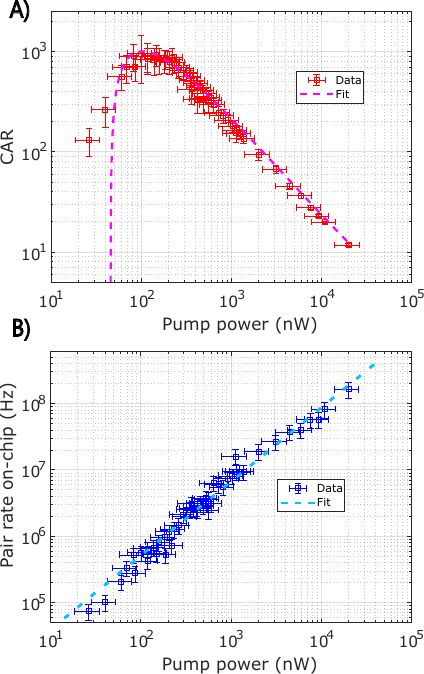}
\caption{A) Coincidence-to-accidentals ratio (CAR) as a function of on-chip pump power. The fitted model is taken from \cite{prabhakar2020two}. B) On-chip pair generation rate as a function of on-chip pump power. This is the rate of pair generation at the output of the poled region of the waveguide. Pump input coupling losses are calculated based on a parametric gain measurement detailed in the supplementary.}
\label{fig:car_data}
\end{figure}

\section{Measurement and Data Processing}

To assess the CAR and on-chip pair generation rate, an HBT experiment (Fig.\ref{fig:exp_setup}B) is performed by passing photon pairs from the chip into a 50/50 beamsplitter and looking for coincidences at the output ports using a pair of superconducting nanowire single photon detectors (SNSPDs) from IDQuantique. True coincidences are created by entangled pairs of photons hitting the detectors at the same time, whereas accidentals are created by two photons from unrelated SNSPD processes causing coincidences. This is a measurement of the signal-to-noise ratio (SNR) of the detection system as the ratio of true to accidental coincidences is determined by the losses in the signal path, the dark count rates for the SNSPDs, and multi-photon generation events. Fig.\ref{fig:raw_hbt}A shows a coincidence histogram collected by the time-to-digital converter recording events from the SNSPDs. The central peak at 0 ns is a result of both true and accidental coincidences, while side peaks are caused by accidentals. For a continuous-wave pump, accidental counts are spread evenly across the delay histogram. For the pulsed case, accidentals counts are localized to multiples of the pump repetition time (4 ns in our case for our 250-MHz repetition rate). The CAR is defined as

\begin{equation} 
CAR = \frac{R_{si}-R_{acc}}{R_{acc}}
\end{equation}

where $R_{si}$ is the total coincidence peak cont rate and $R_{acc}$ is the accidental peak count rate. Error bars are calculated via the standard deviation of CAR calculated from different accidentals peaks. The timing jitter in the SNSPD measurement is limited by the jitter of the electronics (100 ps), resulting in coincidence and accidentals peaks which are much wider than the temporal width of the biphotons themselves. Fig.\ref{fig:raw_hbt}B shows the count rates at the signal and idler detectors as a function of the on-chip pump power. By combining individual detector count rates with the coincidence count rates, the pair generation rate on-chip and detection efficiency for the signal and idler paths can be calculated from fitting the simplified linear model:

\begin{equation}
\label{eq1}
    R_{s} = \epsilon P \eta_{s} + R_{d_{s}}
\end{equation}
\begin{equation}
\label{eq2}
    R_{i} = \epsilon P \eta_{i} + R_{d_{i}}
\end{equation}
\begin{equation}
\label{eq3}
    R_{si} = \frac{1}{2}\epsilon P \eta_{s} \eta_{i}
\end{equation}

where $R_{s,i,si}$ are the signal, idler, and coincidence count rates respectively. $P$ is the pump power in mW, $\epsilon$ is the generation rate in pairs per mW, $\eta_{s,i}$ are the losses for the signal and idler paths including detector efficiency, and $R_{d_{s,i}}$ is the dark count for the signal and idler detectors. The factor of $\frac{1}{2}$ in Eq.4 is the result of a lack of deterministic signal/idler separation in our experiment as they are degenerate in wavelength and polarization. This model is valid for small values of $P$ where the pair generation rate is directly proportional to the input power for SPDC. To compute on-chip power for rate normalization, we measure the input loss by subtracting measurements of the output lensed fiber and throughput losses. The output loss for the lensed fiber is calculated by fitting the parametric generation output power vs pump power as detailed in the supplementary. Using this method, we calculate an output coupling loss of 9 dB from the chip to lensed fiber. The losses for additional components are measured with a 2 \textmu m diode laser. We measure 5 dB from the freespace filter, 3 dB from the transition from SM-2000 to SMF-28 fiber, and 6 dB from the SNSPD's detection efficiency for a total of 23 dB. This agrees well with the measured total system efficiency of 25 $\pm$ 2 dB from fitting experimental data to equations 2, 3, and 4.

%State results for CAR, pair rate, normalized rate, and channel losses. Also talk about input loss measurement
Fig.\ref{fig:car_data}A displays the measured CAR values vs the on-chip pump power which has been fitted with a model provided by \cite{prabhakar2020two}. At high pump powers, the CAR is limited by noise induced from multiphoton events generated by a strong pump pulse. As the pump power is lowered, the CAR increases until the maximum SNR is achieved. The CAR then decreases at lower pump powers due to a loss of signal altogether from the reduced pair generation rate. We measure a maximum CAR of 945 $\pm$ 475 at a pump power of 100 nW. The large variance is a result of the lack of prominent accidental peaks at low pump powers. The theoretical model used to fit the CAR vs power curve depends only on the pair generation rate, detector dark count rates, and system losses. A perfectly achromatic detection system would measure a higher CAR in the absence of the filter due to the higher observed pair generation rate. This is not the case for our system as our SNSPDs and fibers suffer higher losses at longer wavelengths, and therefore a bandpass filter in the measurement path increases our measured CAR by limiting the spectral dependence of the measurement efficiency to help satisfy the assumptions of the model. Fig.\ref{fig:car_data}B displays the measured on-chip pair generation rate as a function of the on-chip power. This is the rate at which generated photon pairs exit the poled region of the waveguide before incurring detection losses. We measure a slope of 8.8 $\pm$ 2.3 GHz/mW over the entire SPDC spectrum. Using our filter bandwidth, this normalizes to a pair-generation rate of 440 $\pm$ 115 kHz/mW/GHz.

\begin{figure}
\centering\includegraphics[width=8cm]{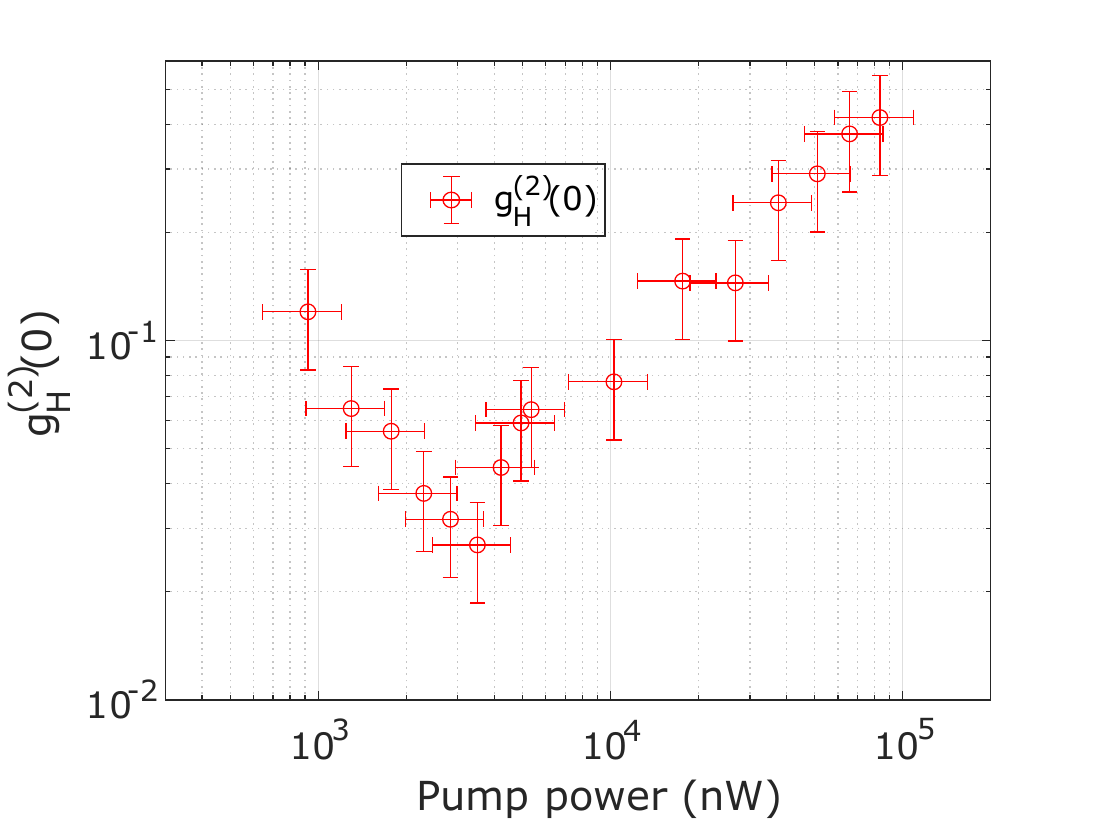}
\caption{Heralded $g_H^{(2)}(0) $ of our source as a function of on-chip pump power.}
\label{fig:heralded_g2}
\end{figure}

In many applications requiring single photons, biphoton sources are used as a heralded source of single photons where the idler is detected to infer the presence of a single photon on the signal channel. To assess the performance of our biphoton source as a heralded single photon source, we measure $g_H^{(2)}(0)$ (Fig.\ref{fig:exp_setup}C). The heralded $g_H^{(2)}(0)$ is defined as

\begin{equation} 
g_H^{(2)}(0) = \frac{R_{HSI}R_{H}}{R_{HS}R_{HI}} = \frac{P_{HSI}}{P_{HS}P_{HI}}
\end{equation}

where $R_{H,S,I}$ are the 2 and 3-fold coincidence rates of the heralding, signal, and idler detectors respectively.  $P_{H,S,I}$ represent the probability for any given laser pulse to produce a click at the subscripted detectors \cite{signorini2020chip}. Fig.\ref{fig:heralded_g2} shows the measured $g_H^{(2)}(0)$ as a function of the on-chip pump power. As the pump power decreases, $g_H^{(2)}(0)$ initially drops as fewer and fewer multiphoton events are observed. At very low pump powers, SNR is lost due to low count rates, causing $g_H^{(2)}(0)$ to rise again. We report a minimum $g_H^{(2)}(0)$ of 0.027$\pm$ 0.0084 at a pump power of 3.5 \textmu W on chip, showing that we can reliably herald single photons from our device using the measurement setup. This performance is competitive with state-of-the-art mid and near-IR sources (see table 1 in the supplementary). The CAR and $g_H^{(2)}(0)$ performance are maximized at two different pump powers because the $g_H^{(2)}(0)$ measurement uses 3 detectors (as opposed to 2 for CAR) and is therefore more susceptible to system losses, necessitating a higher power to maximize the system SNR. Increasing overall system losses would also shift the maximum CAR to a higher power as these losses translate to a lower overall signal.

\begin{figure}
\centering\includegraphics[width=8cm]{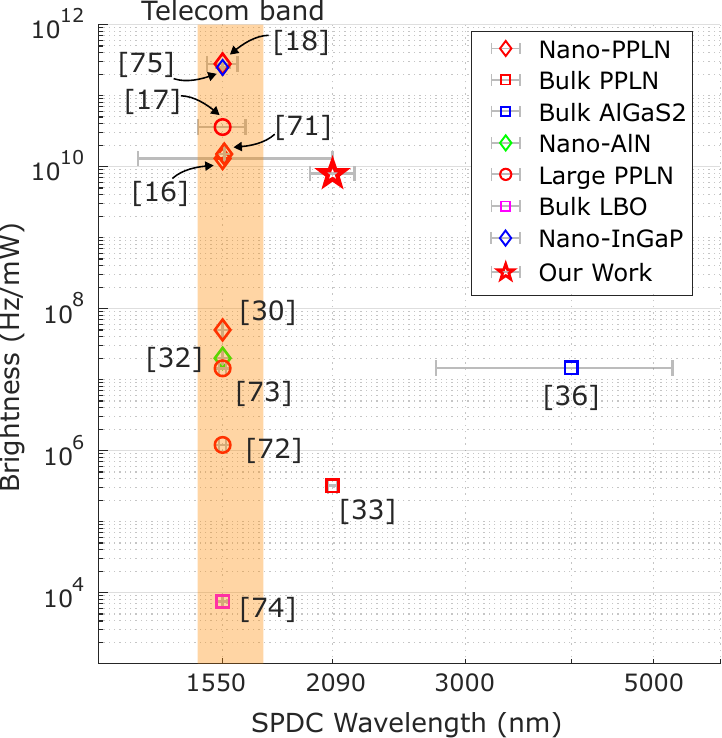}
\caption{A comparison plot of relevant photon pair sources. Horizontal error bars represent the reported source bandwidth. Data for efficiency, wavelength, and bandwidth are taken from \cite{javid2021ultrabroadband, zhao2020high, xue2021ultrabright, bock2016highly} are PPLN ridge waveguides, similar to our device. \cite{zhang2023scalable, yoshizawa2003generation, mori2006distribution} are larger (>1-\textmu m tall) PPLN waveguides. \cite{prabhakar2020two} is a bulk PPLN crystal. \cite{kumar2021mid} is a bulk AlGaS2 crystal. \cite{guo2017parametric} is a nanophotonic AlN ring resonator. \cite{noh2006noncollinear} is a bulk LBO crystal. \cite{zhao2022ingap} is an InGaP ring resonator.}
\label{fig:comp_plot}
\end{figure}

\section{Comparison and Discussion}

Fig.\ref{fig:comp_plot} compares a variety of different IR pair sources with our work. We include 1550-nm sources to demonstrate that not only does our device have state-of-the-art performance at 2 \textmu m, but it is also competitive with state-of-the-art devices for standard telecommunications bands. This allows for a more fair comparison of pair generation rate as most reported 2-\textmu m sources utilize the $\chi^{(3)}$ instead of the $\chi^{(2)}$ nonlinearity, which has a normalized pair generation rate relative to $mW^2$ of pump power instead of $mW$. Table 1 in the supplementary goes into greater detail by normalizing the efficiency with the device length and bandwidth for an apples-to-apples comparison. Looking at the last pair generation rate column ($Hz/mW/cm^2/GHz$), we see that our rate measurements agree well with other devices reported in literature except for one recent demonstration which exceeds the normalized pair generation of all other listed sources by an order of magnitude\cite{xue2021ultrabright}. It should be noted that the pair generation rate will intrinsically be lower at 2 \textmu m compared to 1550 nm as a result of the $\lambda^{-4}$ dependence \cite{jankowski2021dispersion}. In the lower portion of the table which lists 2-\textmu m sources, we see that our source exceeds the CAR, pair generation rate, and heralded $g_H^{(2)}(0)$ of the other recent 2-\textmu m demonstrations in bulk crystals. We attribute this namely to the use of nanophotonic TFLN, which is responsible for the stronger nonlinear interactions given the device length, mode confinement, and dispersion engineering. 

\section{Conclusion}

We have demonstrated a dispersion engineered broadband biphoton source with a record breaking CAR, pair generation rate, and heralded $g_H^{(2)}(0)$ in the 2-\textmu m window. The near-zero GVM and GVD operation of our source allows for use of ultrashort pulses, a crucial resource for realizing large-scale quantum information processing systems as the temporal confinement of these pulses can be used to both enhance the strength of nonlinear interactions as well as create well-defined time bins for temporal multiplexing of quantum gates beyond 1-THz clock speeds. Reduced solar irradiance near 2-\textmu m combined with technological advancements in transmission and detection in the mid-IR make our source an ideal candidate for both freespace and fiber based quantum key distribution. Future work will revolve around characterizing the quantum interference, temporal width, and entanglement properties of the source through Houng-Ou-Mandel (HOM) and Franson interferometry. We plan to improve outcoupling losses via both on-chip adiabatic couplers for pump rejection and the use of an inverse tapered waveguide for better mode matching with a lensed fiber. Using in-house adiabatic coupler designs and state-of-the-art inverse tapers \cite{yao2020efficient}, we can lower the output coupling losses to 0.32 dB and filter losses to 3 dB for a total detection loss of 9.32 dB. 
		
\section{Acknowledgements and Disclosures}

Device nanofabrication was performed at the Kavli Nanoscience Institute (KNI) at Caltech. The authors gratefully acknowledge support from ARO grant no. W911NF-23-1-0048, NSF grant no. 1846273 and 1918549, AFOSR award FA9550-20-1-0040, the center for sensing to intelligence at Caltech, and NASA/JPL. The authors wish to thank NTT Research for their financial support. All authors have accepted responsibility for the entire content of this manuscript and approved its submission. L.L. and A.M. are involved in developing photonic integrated nonlinear circuits at PINC Technologies Inc. L.L. and A.M. have an equity interest in PINC Technologies Inc. The other authors declare that they have no competing interests. Informed consent was obtained from all individuals included in this study. The conducted research is not related to either human or animals use. Data sets used in this paper are available from the corresponding authors upon request.

%Please use the IEEE style for the reference list.
%\bibliographystyle{IEEEtran}
%\bibliography{sample}
\printbibliography

\end{document}